\documentclass[a4paper,aps,pre,floatfix,twocolumn,showpacs]{revtex4}

\usepackage[dvips]{epsfig}
\usepackage[dvips]{graphicx}
\usepackage{latexsym,amsmath,verbatim}
\usepackage{color}

\newcommand{\av}[1]{\langle {#1} \rangle}
\newcommand{\avk}{\left< k \right>}

\begin{document}

\title{Diffusive dynamics on weighted networks}

\title{Mean-field diffusive dynamics on weighted networks} 
\author{Andrea Baronchelli}
\author{Romualdo Pastor-Satorras}
\affiliation{Departament de F\'\i sica i Enginyeria Nuclear, Universitat
  Polit\`ecnica de Catalunya, Campus Nord B4, 08034 Barcelona, Spain}

\date{\today}

\begin{abstract}
  Diffusion is a key element of a large set of phenomena occurring on
  natural and social systems modeled in terms of complex weighted
  networks.  Here, we introduce a general formalism that allows to
  easily write down mean-field equations for any diffusive dynamics on
  weighted networks. We also propose the concept of annealed weighted
  networks, in which such equations become exact. We show the validity
  of our approach addressing the problem of the random walk process,
  pointing out a strong departure of the behavior observed in quenched
  real scale-free networks from the mean-field predictions.
  Additionally, we show how to employ our formalism for more complex
  dynamics.  Our work sheds light on mean-field theory on weighted
  networks and on its range of validity, and warns about the
  reliability of mean-field results for complex dynamics.
\end{abstract}


\maketitle

\section{Introduction}
\label{sec:intro}

Weighted networks represent the natural framework to describe natural,
social, and technological systems, in which the intensity of a
relation or the traffic between elements are important parameters
\cite{Barthelemy:2005,barrat07:_archit}.  In general terms, weighted
networks (WN) are an extension of the concept of network or graph
\cite{Dorogovtsev:2003,Albert:2002}, in which each edge between
vertices $i$ and $j$ has associated a variable, called weight, taking
the form $\omega_{ij} = g(i,j) a_{ij}$, where $a_{ij}$ is the
adjacency matrix and $g(i,j)$ is a function of $i$ and $j$
\cite{newman2004awn}.  Practical realizations of weights in real
networks range from the number of passengers traveling yearly between
two airports in the airport network \cite{Barrat:2004b}, to the
intensity of predator-prey interactions in ecosystems \cite{krause03}
or the traffic measured in packets per unit time between routers in
the Internet \cite{RomusVespasbook}.  A non-weighted network can thus
be understood as a binary network, in which $\omega_{ij} = a_{ij}$,
taking the value $1$ when vertices $i$ and $j$ are connected, and zero
otherwise.  Leaving aside the issue of the proper characterization of
the topological observables associated to networks and their weighted
extensions
\cite{barrat07:_archit,Barrat:2004b,Barthelemy:2005,serrano06:_correl,ahnert:016101,garlaschelli:038701},
a most relevant aspect of WN is the effect of the weight distribution
and structure on the properties of dynamical processes taking place on
top of them
\cite{PhysRevE.65.027103,PhysRevLett.95.098701,wu07:_walks,karsai:036116,0256-307X-22-2-068,colizza07:_invas_thres,ISI:000268327600005,yang:046108}.
These can have an impact in problems ranging from the
information transport in mobile communication networks
\cite{onnela2007sat}, the risks of congestion on Internet
\cite{huberman1997sda}, the disease spreading in the air
transportation network \cite{colizza06:prediction} or biological
issues such as the role of weak interactions in ecological networks
\cite{berlow1999sew}.

In general, the theoretical understanding of dynamical processes on
networks is based at a mesoscopic level on the heterogeneous
mean-field (HMF) approximation and the annealed network approach
\cite{dorogovtsev07:_critic_phenom,barrat08:_dynam}, which assume that
the degree is the only relevant variable characterizing the 
vertices, and that dynamical fluctuations are not relevant. All
dynamical variables can therefore be described in terms of
deterministic rate equations, as a function of time and degree.  Here,
we develop the HMF theory for general local diffusive processes on WN
by extending the annealed network approximation
\cite{dorogovtsev07:_critic_phenom} to the weighted case. We also
define the concept of \textit{annealed weighted network}, in which HMF
theory is exact.  We show the validity of our approach by obtaining an
exact mean-field solution of the paradigmatic random walk process on
WN.  A comparison with numerical simulations allows to set the limits
of validity of HMF on quenched weighted networks, showing also that
it yields in some cases severely incorrect results.  Finally, we
present an example of the application of the annealed weighted network
approximation to more complex dynamics on WN to obtain HMF rate
equations.

Our approach allows for a straightforward mean-field description of
diffusive dynamics on weighted networks, but at the same time calls
for a careful interpretation of mean-field results on quenched
networks, an issue already discussed in the non-weighted case
\cite{Castellano:2006,hong07:_comment,castellano07:_reply,Castellano:2008}.
HMF indeed does not take into account the quenched structure of the
considered network, which however can play a fundamental role inducing
strong deviations even in the apparently safe case in which only
degree dependent weights are present. This should be recalled when
mean-field predictions are to be compared with numerical simulations
or used as a baseline to make predictions on processes on real
weighted networks.

\section{Annealed weighted network approximation}
\label{sec:anne-weight-netw}

\subsection{Annealed network approximation in binary networks}

The HMF theory for dynamical processes on binary complex networks is
ultimately based on the annealed network approach
\cite{dorogovtsev07:_critic_phenom}, that consists in replacing the
adjacency matrix $a_{ij}$, which in a real quenched network is
composed by zeros and ones at fixed positions, by the probability
$\bar{a}_{ij}$ that vertices $i$ and $j$ are connected.
%In a degree
%uncorrelated network, this probability takes the form $\bar{a}_{ij}^0
%= k_i k_j / (\avk N)$ \cite{dorogovtsev07:_critic_phenom}.  
At a statistical degree coarse-grained level, we can replace $a_{ij}$
by the ensemble average $\bar{a}(k_i,k_j)$ \cite{boguna09:_langev},
defined as the probability that two vertices of degree $k_i$ and $k_j$
are connected; that is, by the degree class average (for undirected
networks)
\begin{equation}
  \bar{a}(k, k') = \frac{1}{NP(k)} \frac{1}{N P(k')} \sum_{i\in k}
  \sum_{j\in 
    k'} a_{ij} \equiv \frac{k P(k'|k)}{N P(k')},
  \label{eq:1}
\end{equation}
where $i \in k$ denotes a sum over the set of vertices of degree $k$,
$P(k)$ is the degree distribution (probability that a randomly chosen
vertex is connected with other $k$ vertices) \cite{Albert:2002} and
$P(k'|k)$ is the conditional probability that a vertex of degree $k$
is connected to a vertex of degree $k'$ \cite{Pastor-Satorras:2001}.
In other words, the original network is replaced by a fully connected
graph in which each edge has associated a connection probability
$\bar{a}(k_i,k_j)$ that depends only on the degree of its endpoints.  The
annealed network approach can be directly applied in calculations at
the microscopic (vertex) level or, otherwise, fruitfully extended to a
more intuitive mesoscopic (degree class) level. In the case of
diffusive dynamics, the reasoning runs as follows.  In diffusive
dynamics, particles or interactions move from one vertex to another by
jumping to a randomly chosen nearest neighbor.  Therefore, the
probability that vertex $i$ interacts with vertex $j$, $P(i \to j)$
(the \textit{propagator} of the diffusive interaction), is given by
\begin{equation}
  P(i \to j) = \frac{a_{ij}}{k_i} \equiv \frac{a_{ij}}{\sum_r a_{ir}}. 
\end{equation}
Performing a coarse-graining in degree classes, we can define the
probability that a vertex of degree $k$ interacts with a
vertex of degree $k'$, namely
\begin{equation}
  P(k \to k') = \frac{[N P(k)]^{-1} \sum_{i \in k} \sum_{j \in k'}
    a_{ij}}{[N P(k)]^{-1}  \sum_{i \in k} \sum_{r} a_{ir}} \equiv P(k'|k),
  \label{eq:2}
\end{equation}
where we have used Eq.~(\ref{eq:1}). With this concept in hand,
writing down the HMF rate equations for any diffusive dynamical
process turns out to be straightforward
\cite{dorogovtsev07:_critic_phenom}.

Equation~(\ref{eq:2}) also encodes the idea of an
\textit{annealed network}
\cite{boguna09:_langev,gil05:_optim_disor,weber07:_gener,Castellano:2008,stauffer_annealed2005}.
In opposition to quenched networks, in which edges are frozen between
pairs of vertices, annealed networks are dynamical objects, changing in
time over a certain timescale $\tau_N$, while keeping constant the
degree distribution $P(k)$ and correlations $P(k'|k)$.  When the
network time scale is very small, $\tau_N \to 0$, connections in the
network are completely reshuffled between any two microscopic steps of
the dynamics. In this case, the annealed network is completely defined
by the quantities $P(k)$ and $P(k'|k)$. From a computational point of
view, a diffusive dynamics is easily simulated in these networks, by
assigning to each vertex of degree $k$ nearest neighbors that are
randomly selected among all the vertices in the network, according to
the probability $P(k \to k') = P(k'|k)$.  In the case of degree
uncorrelated networks, this interaction probability simplifies further
to $k' P(k')/ \avk$, so that nearest neighbors are randomly chosen
with probability proportional to their degree. Due to the very
definition of the annealed network approach, HMF theory is exact in
annealed networks \footnote{We do not consider here the case of complex trees,
that would deserve a dedicated analysis due to its specificity  \cite{baronchelli08:_random}.}.

\subsection{Extension to weighted networks}

We now generalize the annealed network approach to the weighted case.
Assuming that the diffusion process is local (depends only on the
departing and arriving vertices), the most general way to take into
account weights in the random walk is to define the probability that a
vertex $i$ interacts with vertex $j$ as a general function of the
weight $\omega_{ij}$. The normalized form of this probability will be
\begin{equation} 
  P_w(i \to j) =  \frac{F[\omega_{ij}]}{\sum_r F[\omega_{ir}]}.
  \label{eq:8}
\end{equation}
The only restriction we impose to the function $F[x]$ is that
$F[0]=0$, in order to avoid diffusion between disconnected vertices
(i.e., with $a_{ij}=0$). This implies that we can write 
\begin{equation}
  F[\omega_{ij}] \equiv F[g(i,j)] a_{ij}.
  \label{eq:4}
\end{equation}
At a mesoscopic level, we again consider the average probability of
interaction of a vertex $k$ with a vertex $k'$, that is defined, in
analogy with Eq.~(\ref{eq:2}), by
\begin{equation}
  P_w(k \to k') = \frac{[N P(k)]^{-1} \sum_{i \in k} \sum_{j \in k'}
    F[\omega_{ij}]}{[N P(k)]^{-1}  \sum_{i \in k} \sum_{r}
    F[\omega_{ir}]} 
  \label{eq:3}
\end{equation}

At the HMF level, where the vertices's degrees are the only
relevant topological variables, this expression can be simplified
considering that the dependence of the weight on the vertices at the
endpoints of each edge can be expressed as a function of the
corresponding degrees, that is, $\omega_{ij} = \bar{g}(k_i, k_j)a_{ij}$.
Using this relation, we have from Eqs.~(\ref{eq:4}) and~(\ref{eq:1})
\begin{widetext}
\begin{eqnarray}
  \frac{\sum_{i \in k} \sum_{j \in k'} F[\omega_{ij}]}{NP(k)}  &=&
  \frac{\sum_{i \in k} \sum_{j \in k'} F[g(i,j)]a_{ij}}{NP(k)}
  = F[\bar{g}(k,k')] \frac{\sum_{i \in k}\sum_{j \in k'} a_{ij}}{N P(k)}
  \equiv k  F[\bar{g}(k,k')]  P(k'|k),\\  
  \frac{\sum_{i \in k} \sum_{r} F[\omega_{ir}]}{N P(k)} &=& 
   \frac{\sum_{i \in k} \sum_{r} F[g(i,j)]a_{ij}}{N P(k)} 
   =  \sum_{q} F[\bar{g}(k,q)] \frac{\sum_{i \in k}\sum_{j \in q} a_{ij}}{N P(k)}
   \equiv  k \sum_{q}  F[\bar{g}(k, q)]  P(q|k). 
\end{eqnarray}
\end{widetext}
From this expressions, we finally obtain the degree coarse-grained
weighted propagator
\begin{eqnarray}
   P_w(k \to k') = \frac{F[\bar{g}(k, k')]  P(k'|k)}{ \sum_q F[\bar{g}(k,q)]
     P(q|k)},  
   \label{eq:6}
\end{eqnarray}
satisfying the normalization condition $\sum_{k'} P_w(k \to k') =1$.

Simpler expressions for the weighted propagator can be obtained if we
consider a linear diffusion process
\cite{colizza07:_invas_thres,wu07:_walks}, proportional to the weight
$\omega_{ij}$, in which $F(x)=x$. Furthermore, considering weights
that are symmetric multiplicative functions of the degrees at the
edges' endpoints, namely $\bar{g}(k,k') \equiv \bar{g}_s(k) \bar{g}_s(k')$ (as is the
case, for example, in the airport transportation network
\cite{Barrat:2004b}), and a degree uncorrelated network, with
$P(k'|k) = k' P(k')/ \avk$, we are led to the simple expression
\begin{equation}
   P_w(k \to k') =\frac{k' \bar{g}_s(k') P(k')}{\av{k \bar{g}_s(k)}},
   \label{eq:10}
\end{equation}
where we define $\av{G(k)} \equiv \sum_k G(k) P(k)$.

Writing the mean-field rate equations for most diffusive dynamical
processes on WN is now easy: The analytics describing the dynamics on
simple binary networks can be generalized to the weighted case simply
by replacing the propagator $P(k \to k')$ with $P_w(k \to k')$.

The spirit in which the HMF rate equations are constructed leads also
to introduce the concept of an \textit{annealed weighted
  network}.  In this case, in the event of an interaction, a vertex of
degree $k$ chooses as the interaction target a vertex of degree $k'$,
randomly selected among all vertices in the network with probability
$P_w(k \to k')$. Therefore, an annealed weighted network can be
understood as an annealed binary network, which is completely
reshuffled between dynamic time steps, preserving its degree
distribution $P(k)$ and an effective degree correlation pattern given
by $P_{\mathrm{eff}}(k'|k) \equiv P_w(k \to k')$.  Again, in the case
of linear diffusion in a degree uncorrelated network with symmetric
multiplicative weights, this connection probability simplifies to $k'
\bar{g}_s(k') P(k')/ \av{k \bar{g}_s(k)}$, so that the interacting vertex is
chosen with probability proportional to $k \bar{g}_s(k)$. Simulations run on
such networks are described exactly by weighted HMF rate
equations. Quenched WN can in principle determine different behaviors,
and, as we will see below, this is in fact the case.

\section{Random walks on weighted networks}
\label{sec:random-walk-weighted}

\subsection{Weighted heterogeneous mean-field solution }
\label{sec:anne-weight-netw-1}

To check the application of the formalism derived above, let us
consider as an example the simple yet paradigmatic random walk
process  \cite{noh04:_random_walks_compl_networ}. This is defined by
a walker that, located on a given vertex of degree $k$ at time $t$,
hops to one of the $k$ neighbors of that vertex at time $t + 1$,
randomly chosen with probability depending on the weights connecting
the vertices. The simplicity of this problem allows to solve it
exactly for certain forms of the weight structure
\cite{wu07:_walks,fronczak:016107}, following the master equation
approach developed in
Ref.~\cite{noh04:_random_walks_compl_networ}. Here we focus on the
application of the simpler HFM in the most general case, which
provides results for any weight pattern.

To construct the appropriate HMF rate equation, we consider the
probability $R_k(t)$ that the walker is in \textit{any} vertex of
degree $k$, which fulfills the master equation
\begin{eqnarray}
  \dot{R}_k(t)  &=& - \sum_{k'}   R_k(t)  P_w(k \to k') + \sum_{k'}
  P_w(k' \to k) R_{k'}(t) \nonumber \\
  &=& R_k(t) + \sum_{k'} P_w(k' \to k) R_{k'}(t).
  \label{eq:444}
\end{eqnarray}
The first term in Eq.~(\ref{eq:444}) represents the outflow of
probability due to walkers abandoning vertices of degree $k$, while
the second term represents walkers arriving to vertices of degree $k$
from vertices of degree $k'$, following a weighted random step.
In the steady state this equation takes the iterative form
\begin{equation}
  R_k = \sum_{k'} P_w(k' \to k) R_{k'}.
  \label{eq:14}
\end{equation}
Interpreted in terms of a Markov process \cite{durret99:_essen},
the steady state solution is given, for any degree correlation and
weight pattern, by the expression
\begin{equation}
  R_k = \lim_{n\to\infty} P^{(n)}_w(k \to k),
  \label{eq:15}
\end{equation}
where $P^{(n)}_w$ is the $n$-th power of the probability $P_w(k \to
k')$, considered as a matrix.  Explicit solutions in the steady state
can be found for some weight patterns by imposing the detailed balance
condition $R_k P_w(k \to k') = R_{k'} P_w(k' \to k)$
\cite{durret99:_essen}, which ensures a solution of
Eq.~(\ref{eq:14}). From this condition we obtain, for a general
diffusion process,
\begin{eqnarray}
  \frac{R_k}{R_{k'}} &=& \frac{ P_w(k' \to k)}{ P_w(k \to k')}
  \nonumber \\ 
  &\equiv&
  \frac{F[\bar{g}(k',k)] k P(k)}{F[\bar{g}(k,k')] k' P(k')} \frac{ \sum_q
    F[\bar{g}(k,q)] P(q|k)} 
  {\sum_q F[\bar{g}(k',q)]P(q|k')},
  \label{eq:11}
\end{eqnarray}
% \begin{widetext}
% \begin{equation}
%   \frac{R_k}{R_{k'}} = \frac{ P_w(k' \to k)}{ P_w(k \to k')} 
%   \equiv
%   \frac{F[\bar{g}(k',k)] k P(k)}{F[\bar{g}(k,k')] k' P(k')} \frac{ \sum_q
%     F[\bar{g}(k,q)] P(q|k)} 
%   {\sum_q F[\bar{g}(k',q)]P(q|k')},
%   \label{eq:11}
% \end{equation}
% \end{widetext}
where in the last equality we have used the degree detailed balance
condition $k' P(k') P(k|k') = k P(k) P(k'|k)$ \cite{Boguna:2002}.
Then, for symmetric weights $\omega_{ij}=\omega_{ji}$, or, at the
degree level, $\bar{g}(k,k') \equiv \bar{g}_s(k,k') = \bar{g}_s(k',k)$, we obtain the
normalized probability
\begin{equation}f
  R_k =  \frac{k P(k) \sum_q F[\bar{g}_s(k,q)] P(q|k)}{ \sum_{k'} \sum_q k'
    P(k') F[\bar{g}_s(k',q)]  P(q|k')}.
  \label{eq:12}
\end{equation}
If the weights only depend on the first vertex in the edge, that is,
$\omega_{ij} = g_1(i) a_{ij}$ or $\bar{g}(k, k') \equiv \bar{g}_1(k)$, the effect
of weights effectively vanishes and we recover diffusion in a binary
network \cite{noh04:_random_walks_compl_networ},
\begin{equation}
  R_k =  \frac{k P(k)}{\sum_{k'} k' P(k')}
\end{equation}
Finally, if the weights only depend on the second vertex in the edge,
that is, $\omega_{ij} = g_2(j) a_{ij}$ or $\bar{g}(k, k') \equiv \bar{g}_2(k')$,
then we obtain
\begin{equation}
  R_k = \frac{k P(k) F[\bar{g}_2(k)] \sum_q F[\bar{g}_2(q)] P(q|k)}{ \sum_{k'} \sum_q k'
    P(k') F[\bar{g}_2(k')]F[\bar{g}_2(q)]  P(q|k')}.
  \label{eq:13}
\end{equation}
Noticeably, results (\ref{eq:12}) and (\ref{eq:13}) coincide if
diffusion is linear, $F(x)=x$, and the weights are symmetric and
multiplicative, $\bar{g}_s(k,k') = \bar{g}_s(k) \bar{g}_s(k')$, recovering the results
reported in \cite{wu07:_walks,fronczak:016107}.

Further quantities can be analogously computed with ease in the
annealed WN approximation. For example, the coverage $S(t)$
\cite{montroll:167,stauffer_annealed2005,baronchelli08:_random},
defined as the average number of different vertices visited by the
walker at time $t$, can be computed as follows. Let us define
$s_k(t)=v_k(t)/[NP(k)]$ as the fraction of vertices of degree $k$
visited by the random walker, $v_k(t)$ being total number of such
vertices. From here, we have $S(t) = N \sum_k P(k) s_{k}(t)$. The
quantity $v_k(t)$ increases in time as the random walk arrives to
vertices that have never been visited. Therefore, at a mean-field
level, it fulfills the rate equation \cite{baronchelli08:_random}
\begin{equation}
  \dot{v}_k(t) = [1-s_k(t)] \sum_{k'} P_w(k' \to k) R_{k'}(t).
\end{equation}
Dividing this equation by $N P(k)$ and using Eq.~(\ref{eq:6}), we are
led in the general case to 
\begin{widetext}
  \begin{equation}
  \dot{s}_k(t) = [1-s_k(t)]k 
  \sum_{k'} \frac{F[\bar{g}(k',k)] P(k'|k)}{N k'
    P(k')\sum_q  F[\bar{g}(k',q)]
    P(q|k')} R_{k'}(t).
  \label{eq:7}
\end{equation}
\end{widetext}
Assuming now that $R_k(t)$ reaches its steady state in a very short
time, of order $1$, we can substitute the steady state value $R_k$
into Eq.~(\ref{eq:7}) and integrate it, with the initial condition
$s_k(0)=0$. For symmetric weights, Eq.~(\ref{eq:12}) the integrations
leads to the result
\begin{widetext}
\begin{equation}
  s_k(t) = 1-\exp\left\{ -t \; \frac{ k  \sum_{k'} F[\bar{g}_s(k',k)]
      P(k'|k)}{N 
      \sum_{k''}\sum_q k'' P(k'') F[\bar{g}_s(k'', q)] P(q|k'')}\right\}.  
  \label{eq:9}
\end{equation}
\end{widetext}
Moreover, straightforward mean-field arguments
\cite{saramaki2004sfn,baronchelli2003rsa,baronchelli08:_random}
provide an expression for the mean first-passage time (MFPT) $\tau_k$,
defined as the average time that a walker takes to arrive to a given
vertex of degree $k$, starting from a randomly chosen initial vertex
\cite{noh04:_random_walks_compl_networ}. Let us define the occupation
probability $\rho_k(t) = R_k(t)/ [N P(k)]$ as the probability that the
walker is at a \textit{given} vertex of degree $k$. Thus, the
probability for the walker to arrive at a vertex $i$, in a hop
following a randomly chosen weighted edge, is given by $q(i) =
\rho_{k_i}(t)$. Therefore, the probability of arriving at vertex $i$
for the first time after $t$ hops is $P_a(i;t) = [1-q(i)]^{t-1}
q(i)$. The MFPT to vertex $i$ can thus be estimated as the average
\begin{equation}
  \tau_{k_i} = \sum_t t P_a(i;t) = \frac{1}{q(i)}.
\end{equation}
We have thus that the MFPT is proportional to the inverse of the steady
state probability, $R_k$, namely
\begin{equation}
  \tau_k = \frac{N P(k)}{R_k}.
\end{equation}
This simple HMF approximation can be improved by the inclusion of
sub-leading terms using more formal techniques
\cite{noh04:_random_walks_compl_networ,fronczak:016107}.

\subsection{Dynamics on quenched weighted networks}
\label{sec:dynam-quench-weight}

The weighted HMF expressions obtained above will exactly match, by
definition, the results of simulations on annealed WN. A different
issue is, however, their validity for real quenched WN, in particular
for networks with a scale-free degree distribution, $P(k) \sim
k^{-\gamma}$ \cite{Dorogovtsev:2003,Albert:2002}. The question of the
non mean field behavior of dynamical processes on binary networks has
been already pointed out in the literature
\cite{Castellano:2006,1742-5468-2006-05-P05001}. As we will see, the
situation can get even worse in the case of WN.

We consider as a particular example the case of liner diffusion on
symmetric multiplicative weighted networks, with weight intensity
$\bar{g}_s(k) = k^\theta $, typical of some real systems such as the airport
network \cite{Barrat:2004b}.  In the limit of large positive
$\theta$, in an annealed WN we would expect the random walker to reach
the largest hub in the first time step, and then stay there forever
\footnote{Remember that in an annealed network the interaction of a
  vertex $i$ with itself is allowed.}.  In a quenched WN, on the other
hand, the walker is expected to commute forever between the first
encountered pair of vertices who have the property of being
reciprocally the highest degree neighbor of each other. In order to
explore quantitatively this intuitive disagreement, we have performed
numerical simulations of random walks in degree uncorrelated and
multiplicative scale-free WN, with weight intensity $\bar{g}_s(k) = k^\theta
$ and linear diffusion, which yield the HMF results
\begin{eqnarray}
  &&\tau_k  = \frac{1}{\rho_k}  =
  \frac{N\av{k^{1+\theta}}}{k^{1+\theta}},\label{eq:mfpt} \\  
  &&1-\frac{S(t)}{N} =  \mathcal{F}_{\gamma, \theta}  \left( \frac{t}{N
      \av{k^{1+\theta}}}  \right) 
  \label{eq:coverage},
\end{eqnarray}
where $\mathcal{F}_{\gamma, \theta}(x)$ is a scaling function depending on
$\gamma$ and $\theta$.  Uncorrelated scale-free quenched networks with
any degree exponent $\gamma$ are created using the Uncorrelated
Configuration Model (UCM)~\cite{Catanzaro:2005}, characterized by a
hard cut-off $k_c=\sqrt{N}$, preventing the generation of correlations
for $\gamma<3$. Random walks are performed by moving the walker from
its present position $i$ to a nearest neighbor $j$, chosen with probability
\begin{equation}
  P(i \to j) = \frac{k_j^\theta a_{ij}}{\sum_q k_q^\theta a_{iq}}.
\end{equation}
%proportional to $\omega_{ij} = (k_i k_j)^\theta
%a_{ij}$. 
On annealed WN, on the other hand, the next step of the walk
is a vertex $j$, randomly chosen among all vertices in the network
with probability proportional to $k_j^{1+\theta}$, see
Eq.~(\ref{eq:10}). In our simulations we keep fixed a degree exponent
$\gamma=2.5$ and a minimum degree of the network $m=4$.

\begin{figure}[!t]
\begin{center}
\includegraphics*[width=8cm]{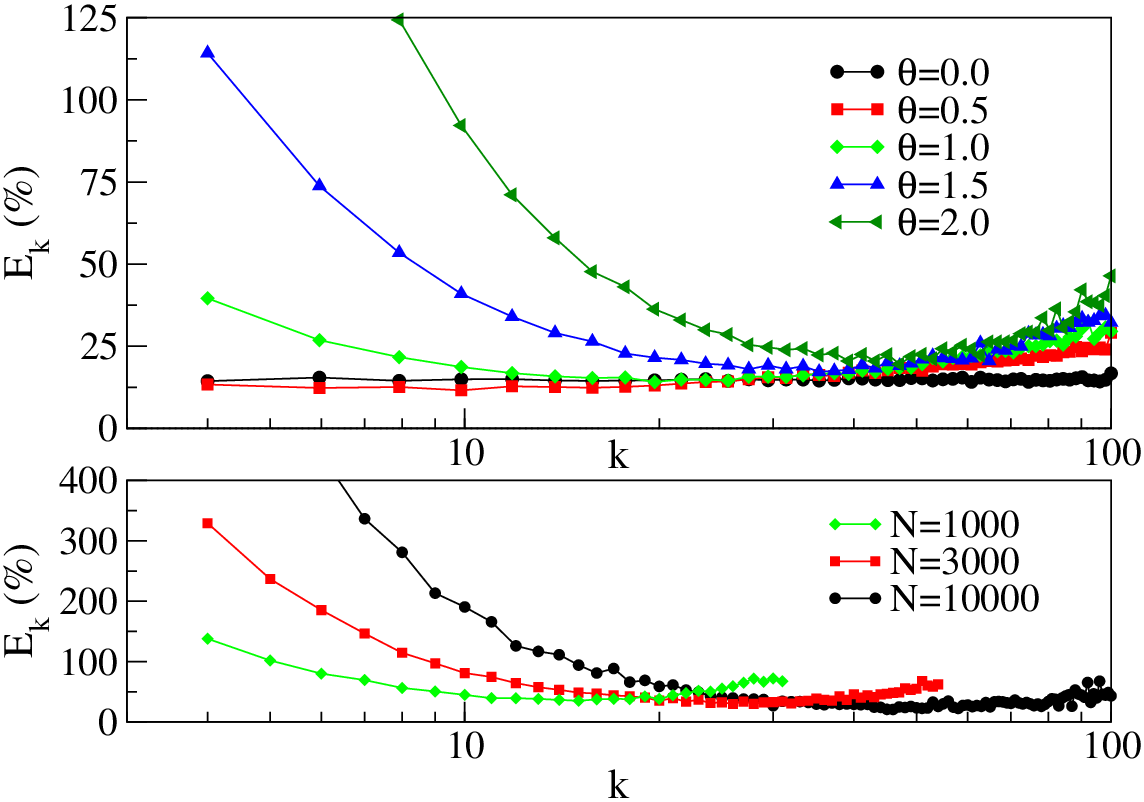}
\end{center}
\caption{(Color Online) Top: Relative error between the theoretical HMF prediction
  Eq.~(\ref{eq:mfpt}) and numerical simulations for the MFPT in
  quenched scale-free WN of size $N=10^4$ and $\gamma = 2.5$.  Bottom: Relative error for
  $\theta = 2.5$ and different network sizes.}
\label{fig:mfpt}
\end{figure}

We first study the validity of the weighted HMF prediction for the
MFPT, by examining the differences between the theoretical expression
$\tau_k^{th}$, Eq.~(\ref{eq:mfpt}), and the numerical result,
$\tau_k^{num}$. 
In particular, we focus on the relative error as a
function of the degree, defined as 
\begin{equation}
  E_k=\frac{\tau_k^{num} -  \tau_k^{th}}{\tau_k^{th}}.
\end{equation}
Fig.~\ref{fig:mfpt}(top) shows the numerical evaluation of this
function in quenched scale-free WN of fixed size $N=10^4$, for
different values of $\theta$ \footnote{Of course, in numerical simulations 
the ensemble average of Eqs. (\ref{eq:mfpt}) is substituted with the 
network average performed on the generated networks.}.  For $\theta=0$, corresponding to a
binary network, we recover a constant baseline error of about $15\%$,
as already reported in the literature \cite{baronchelli2003rsa}.  As
$\theta$ increases, the numerical results start to deviate from the
HMF prediction, the error being larger for small degrees and also
increasing for large $k$. The error in the estimate of the weighted
HMF theory also increases when increasing the system size, as can be
seen from Fig.~\ref{fig:mfpt}(bottom).

\begin{figure}[!t]
\begin{center}
\includegraphics*[width=6.8cm]{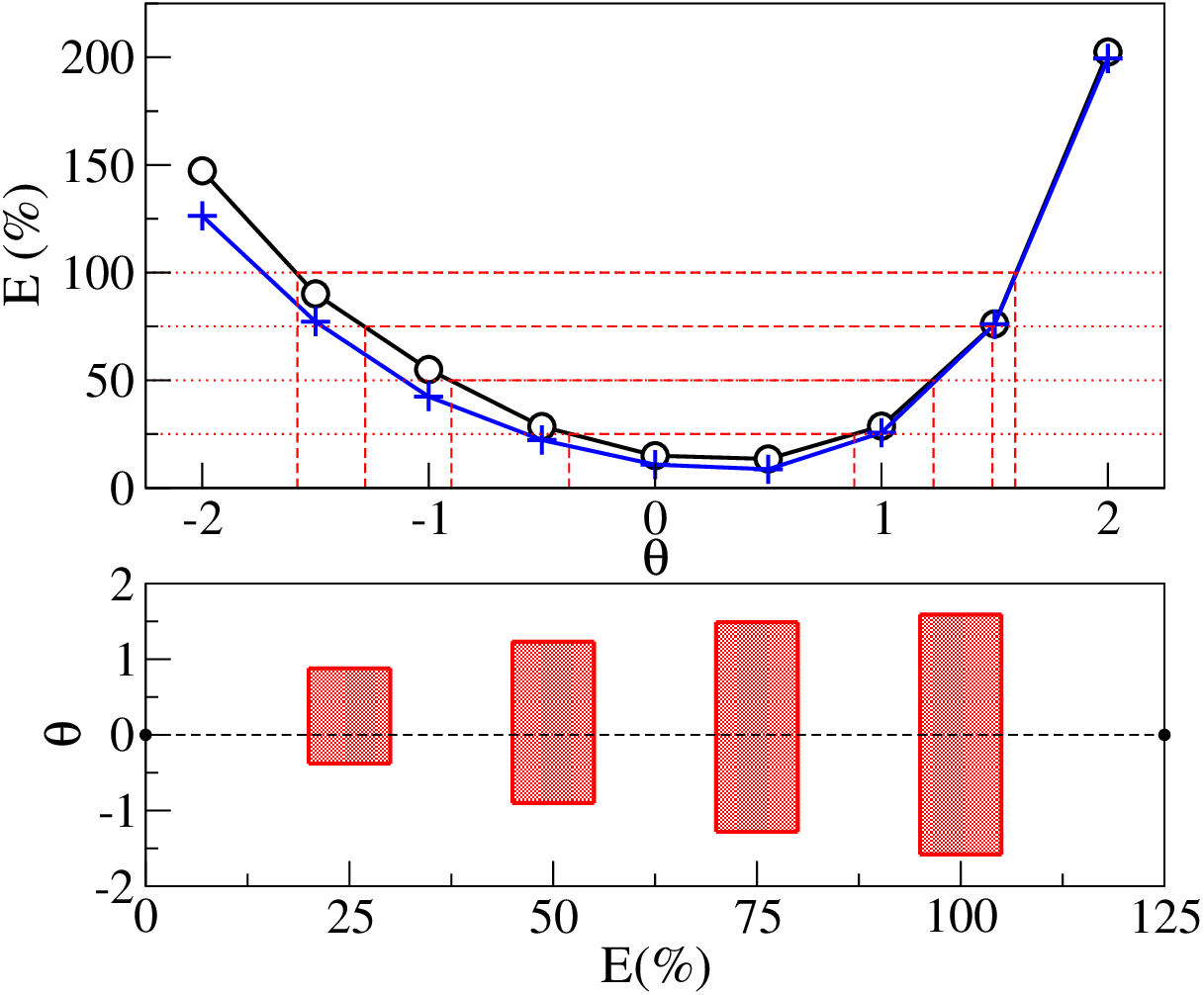}
\end{center}
\caption{(Color Online) Top: Average error as a function of $\theta$ in quenched WN
  of size $N=10^4$ and $\gamma=2.5$ for the simple HMF estimate (circles),
  Eq.~(\ref{eq:mfpt}), and a more refined approximation (crosses),
  Eq.~(\ref{eq:17}). Bottom: Range of $\theta$ values yielding an
  average error $E$ smaller than a given value for the HMF estimate.}
\label{fig:mfpterror}
\end{figure}

In order to provide an estimate of the range of validity of the HMF
approximation, in Fig.~\ref{fig:mfpterror} we explore the average
error as a function of $\theta$, $E = \sum_k P(k) E_k$. As we can see
from Fig.~\ref{fig:mfpterror}(top), for the particular value
$\gamma=2.5$ considered, the error is minimal, compatible with the
baseline of $15\%$, in the interval $\theta \in [0,0.5]$, while it
increases for values of $\theta$ outside this interval. In
Fig.~\ref{fig:mfpterror}(top) we additionally compare the performance
of the HMF prediction with a more elaborate estimate of the MFPT
obtained by the master equation technique, namely
\cite{fronczak:016107}
\begin{equation}
  \tau_k' \simeq
  \frac{N\av{k^{1+\theta}}}{k^{1+\theta}} +
  \frac{N\av{k}^2}{k \av{k^2}} -2.
  \label{eq:17}
\end{equation}
As we can see, Eq.~(\ref{eq:17}) fares slightly better for $\theta<0$,
but the HMF result remains a fairly good first order approximation.

This kind of plot can be used to estimate the range of values of
$\theta$ for which the weighted HMF approximation provides a correct
result within an accepted maximum tolerable error, see
Fig.~\ref{fig:mfpterror}(bottom). Thus, for example, a deviation
smaller than a $50\%$ in scale-free networks with $\gamma=2.5$ is
achieved for values of $\theta \in [-0.9, 1.2]$. Outside this range,
other more sophisticated approaches should be followed.
%, for example the
%master equation approach used in
%\cite{noh04:_random_walks_compl_networ,wu07:_walks,fronczak:016107} in
%the case of the random walk problem.

\begin{figure}[t]
\begin{center}
\includegraphics*[width=8cm]{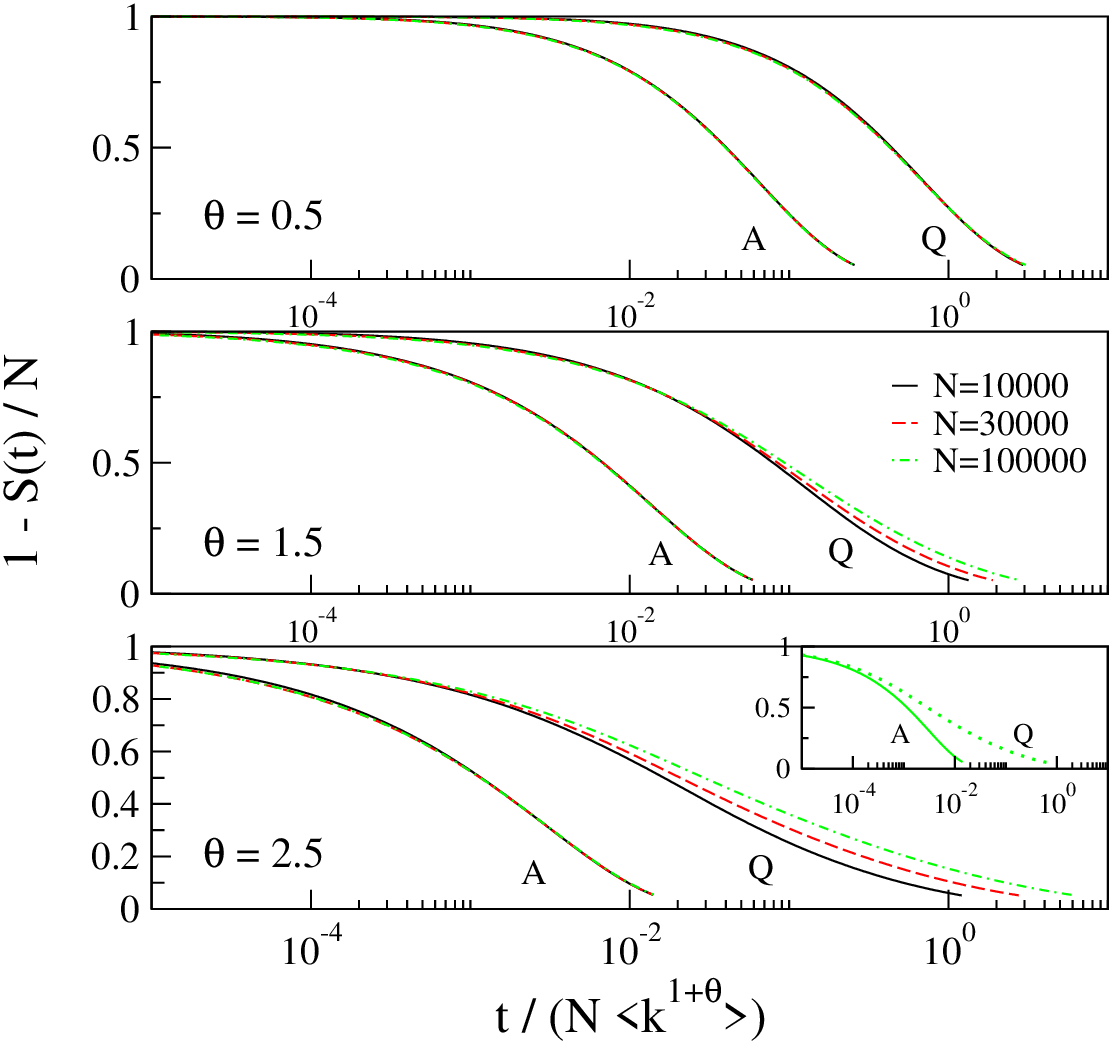}
\end{center}
\caption{(Color Online) Data collapse of the network coverage, as given by
  Eq.~(\ref{eq:coverage}), in annealed ($A$) and quenched ($Q$) scale-free
  WN, with $\gamma=2.5$ for different values of $\theta$ and network sizes.  Curves
  shifted on the horizontal axis for clarity. Inset: Unshifted curves
  for $\theta=2.5$ and $N=10^5$ for annealed
  ($A$, continuous line) and quenched ($Q$, dotted line) WN.}
\label{fig:coverage}
\end{figure}

Concerning the results for the network coverage $S(t)$, a direct
comparison with Eq.~(\ref{eq:coverage}) is not feasible, since the
exact form of the function $\mathcal{F}_{\gamma, \theta}(x)$ depends
on several approximations (steady state approximation in
Eq.~(\ref{eq:9}), continuous degree approximations, etc.). We
therefore take as the main weighted HMF prediction the scaling form in
Eq.~(\ref{eq:coverage}), which we check by means of a data collapse
analysis: If Eq.~(\ref{eq:coverage}) is correct, we expect that plots
of $1-S(t)/N$ as a function of $t/(N \av{k^{1+\theta}})$ will collapse
onto the universal function $\mathcal{F}_{\gamma,\theta}(x)$ when
plotted for different values of N.  In Fig.~\ref{fig:coverage} we
compare the results of simulations on quenched weighted networks with
simulations in the corresponding annealed ones, for different values
of $\theta$.  For small values of $\theta$ (i.e. smaller than $0.5$),
we observe a perfect agreement in both sets of simulations with the
weighted HMF prediction.  Again, however, the perfect collapse of the
annealed case contrasts with the large deviations shown by quenched
networks for large values of $\theta$, outside the regime of validity
of the weighted annealed network approximation.

The reason for the failure of weighted HMF theory in the random walk
case is easy to understand. For positive $\theta$, a quenched topology
reduces the accessibility of the network by trapping the walker in
pairs of adjacent high degree vertices nodes and, at the same time,
slows down the exploration of the network due to the trapping effect
of high degree vertices \cite{carmi:066111}. These effects, which are
stronger for large $\theta$, explain the large MFPT observed for small
and large degree, Fig.~\ref{fig:mfpt}(top), as well as the larger
times needed to reach a fixed coverage, Fig.~\ref{fig:coverage}.  For
negative $\theta$, on the other hand, the walker is biased towards
small degree vertices, missing thus the hubs that provide connectivity
to far away regions in the network \cite{adamic01}.

\section{General dynamical processes}
\label{sec:gener-dynam-weight}

As we have pointed out in Sec.~\ref{sec:anne-weight-netw}, the
annealed weighted network approximation can be easily applied to
general diffusive processes in weighted networks to obtain an
approximate HMF solution. Within this framework, the weighted HMF rate
equations are obtained from the ones on binary networks by
substituting the binary propagator by the weighted one $P_w(k \to
k')$, Eq.~(\ref{eq:6}).

As a case example of application, we will consider here the case of
the contact process (CP) \cite{Marro}, whose dynamics on a network is
defined as follows \cite{Castellano:2006}: An initial fraction
$\rho_0$ of vertices is randomly chosen and occupied by a particle.
The time evolution of the process runs as follows: At each time step
$t$, a particle in a vertex $i$ is chosen at random. With probability
$p$, this particle disappears.  With probability $1-p$, on the other
hand, the particle may generate an offspring. To do so, a vertex $j$,
nearest neighbor of the vertex $i$, is chosen.  If vertex $j$ is
occupied by a particle, nothing happens; if it is empty, a new
particle is created on $j$.  In any case, time is updated as $t \to
t+1/n(t)$, where $n(t)$ is the number of particles present at the
beginning of the time step. In a binary network, the second vertex $j$
is chosen uniformly at random among all neighbors of $i$, i.e. with
probability $1/k_i$. In a weighted network, on the other hand, it is
selected with probability depending on $\omega_{ij}$.

The key point in the HMF analysis of the CP is the probability
$\rho_k(t)$ that a vertex of degree $k$ is occupied by a particle at
time $t$ \cite{Castellano:2006,Castellano:2008}. To write the HMF
equation for this quantity it is useful to focus instead on the number
of occupied vertices of degree $k$, $n_k(t)$, which is related to
$\rho_k(t)$ by
\begin{equation}
  \rho_k(t) = \frac{n_k(t)}{NP(k)}.
\end{equation}
The rate equation for $n_k(t)$ is easy to obtain by means of
mean-field arguments \cite{Castellano:2006,marianproc}, leading to
\begin{equation}
  \dot{n}_k(t) = -p n_k(t) + (1-p) [1-\rho_k(t)] \sum_{k'} P_w(k' \to
  k) n_{k'}(t).
  \label{eq:5}
\end{equation}
The first term comes from particles disappearing from vertices of
degree $k$ with probability $p$; the second term correspond to the
offsprings of particles in connected vertices of degree $k'$ (with
probability $1-p$), arriving to vertices of degree $k$ which are empty
(with probability $1-\rho_k(t)$). Dividing Eq.~(\ref{eq:5}) by $N
P(k)$, substituting the expression of the weighted propagator
Eq.~(\ref{eq:6}) and applying the degree detailed balance condition
\cite{Boguna:2002} we obtain the general weighted HMF equation for
the CP
\begin{widetext}
\begin{equation}
  \dot{\rho}_k(t) = - \rho_k(t) + \lambda k [1-\rho_k(t)] \sum_{k'}
  \frac{F[\bar{g}(k',k)] P(k'|k)}{k' \sum_q F[\bar{g}(k',q)] P(q|k')} \rho_{k'}(t).
  \label{eq:16}
\end{equation}
\end{widetext}
In this last equation, we have performed a rescaling of time, and
defined the parameter $\lambda= (1-p)/p$. For symmetric multiplicative
weights of the form $\bar{g}(k, k') = \bar{g}_s(k)\bar{g}_s(k')$, and assuming linear
diffusion a degree uncorrelated network, we are led to the simplified
equation
\begin{equation}
 \dot{\rho}_k(t) = - \rho_k(t) + \lambda  [1-\rho_k(t)] \frac{k \bar{g}_s(k)
   \rho(t)}{\av{k \bar{g}_s(k)}},
\end{equation}
where $\rho(t) = \sum_k P(k) \rho_k(t)$. In this expression we recover
the result presented in Ref.~\cite{karsai:036116}.

Along the same lines, more complex dynamical processes subject to
general weighted diffusive rules can also be considered and the
corresponding weighted HMF theories easily developed.

\section{Conclusions}
\label{sec:conclusions}

In this paper we have developed a general formalism to write down HMF
equations for diffusive dynamical processes on WN.  Moreover, we have
introduced the concept of annealed weighted networks, for which HMF
theory represents an exact description.  Considering as a simple
example the random walk process, we have presented exact mean-field
implicit solutions for its behavior for any correlation and weight
pattern, as well as explicit formulas for the case of symmetric
weights. By means of numerical simulations, we have also shown that
weighted HMF theory can describe diffusion in real quenched scale-free
networks with multiplicative weights proportional to $k^\theta$ only
for small values of $|\theta|$. Finally, we have demonstrated how the
annealed weighted approximation can be fruitfully applied to obtain
information for more complex dynamical systems, as for example the
contact process, in which more exact analytical alternatives can be
difficult (or impossible) to work out.

Overall, our work provides a straightforward method to describe any
dynamical process on weighted network in therms of HMF, but it also
puts a word of warning in a \textit{tout court} extrapolation of HMF
results to real quenched weighted network. In fact, while HMF provides
a first estimate of the behavior of diffusive systems when the weights
are not excessively strong, it fails in the regime of large
weights. These observations add a new ingredient to the issue of the
non mean-field behavior observed in several dynamical processes on
binary networks
\cite{Castellano:2006,Castellano:2008,bancal-2009}. This mean-field
failure can in some cases be attributed to the build up of dynamical
correlations between vertices \cite{boguna09:_langev,bancal-2009},
especially at low particle densities, that invalidate
the HMF assumptions. In fact, when considering the effects of weights on
dynamics the situation becomes more complex, as the random walk
analysis proves. In this case, dynamical correlations cannot play any
role, since we are dealing with a single particle. The lack of
mean-field behavior must thus be ascribed to the presence of
\textit{topological traps}, which slow down the dynamics and are
enhanced by the presence of strong weights.  The development of new
theoretical tools beyond mean-field to tackle the understanding of
dynamical systems in such critical structures, and incorporating these
elements, becomes therefore an important future research venue.

\section*{Acknowledgments}

We acknowledge financial support from the
Spanish MEC (FEDER), under project No. FIS2007-66485-C02-01, as well
as additional support through ICREA Academia, funded by the
Generalitat de Catalunya.  A. B.  acknowledges support of Spanish MCI
through the Juan de la Cierva program funded by the European Social
Fund.  We thank the hospitality of the ISI Foundation (Turin, Italy),
where part of this work was developed.

%\bibliographystyle{prsty}
%\bibliography{ref.bib}

\end{document}